\documentstyle[preprint,aps,eqsecnum]{revtex}

\newcommand\beq{\begin{equation}}
\newcommand\eeq{\end{equation}}
\newcommand\bea{\begin{eqnarray}}
\newcommand\eea{\end{eqnarray}}
\newcommand\tr{{\mbox{tr\,}}}

\begin{document}
%%%%%%%%%%%%%%%%
\tighten

\title{Stroboscopic quantization of autonomous systems}
\author{Bruno Eckhardt and Uzy Smilansky 
\footnote{Permanent address:
Department of
Physics of Complex Systems,  The Weizmann Institute of Science, Rehovot
76100, Israel}\\[5mm]}
\address{Fachbereich Physik, Philipps-Universit\"at Marburg,
D-35032 Marburg, Germany
\footnote{The RevTex-files for this article area available from the
first author,\\ email to: bruno.eckhardt@physik.uni-marburg.de}\\[2mm]}
\date{\today}
\maketitle

\begin{abstract}
We introduce a  semiclassical quantization method which is based on
a stroboscopic description of the classical and the quantum flows.
We show that this approach emerges naturally when one is interested
in extracting the energy spectrum within a prescribed and 
{\it finite} energy interval. The resulting semiclassical expression 
involves a finite number of periodic orbits whose energies are in 
the considered interval. Higher order corrections which reflect 
the sharp restriction of the spectrum to an interval are 
explicitely given. The relation to Fourier methods for extracting
semiclassical spectra, such as harmonic inversion, is worked
out. The constraints due to the finite
dimension of the  Hilbert space and the unitarity of the restricted quantum
evolution operator are important ingredients in this context.
\end{abstract}
%%%%%%%%%%%%%%%%%%%%%%%%%%%%%%%%%%%%%%%%%%%%%%%%%%%%%%%%%%%%%%%%%%%%%%%%%%%%%%%%
\draft
\narrowtext

\newpage
\section {Introduction}
\label {section:Introduction}
  The pioneering semiclassical quantization schemes introduced by
Gutzwiller \cite {Gut0,Gutz}
(general chaotic  hamiltonian flows) and  Balian and Bloch (billiards)
\cite{BB70} are afflicted by a severe and intrisic problem: the 
semiclassical trace formulae and the spectral determinants derived from
them involve an unlimited number of periodic orbits and do not converge 
on the real energy axis. Several authors proposed various solutions
to this problem:

The cycle expansion\cite{CE89,Varenna,Pinball} 
exploits the presence of correlations between long  
and short orbits to argue that a suitable rearrangement of the spectral
 determinant converges on the real axis. The improved convergence 
 has been demonstrated for
 certain open systems like scattering off three disks\cite{Pinball,Wirzba} 
 and for a map, 
 the quantum bakers map\cite{EH}. 
 Applications to bounded systems are typically 
 troubled by the lack of a convenient symbolic coding and the presence of 
 marginally stable periodic orbits that spoil the ideal exponential 
 convergence\cite{Tanner,Dahlqvist}. 

 Keating and Berry \cite{BK90,BK92,K92} made use of the  analytic
properties of the exact spectral
determinant, and enforced it on the semiclassical expression, thus deriving
a quantization scheme which is similar to the Riemann-Siegel formula 
for the Riemann $\zeta$ function on the critical line. Its
main term involves a finite number of periodic orbits, and the further
corrections are given explicitly and they are small in the 
semiclassical limit.

 There are, however, alternative  semiclassical  quantization schemes for
chaotic systems: One considers the quantum analogue  of the classical 
evolution operator on a Poincar\'e surface of section of a finite 
volume\cite{Bogomol,DoronUS,USLeshouches}. 
The area preserving Poincar\'e map when quantized yields a
semiclassically unitary evolution operator which acts on a Hilbert space
of dimension $L=\left[ {{\cal V}\over \hbar^f}\right ]$, where  
${\cal V}$ is the phase space volume of the $2f$ dimensional section,
and $[\cdot ]$ stands for the integer part. Denoting by
$S(E,\hbar)$ the $L\times L$ (semiclassically) unitary evolution operator,
one expresses the quantization condition in terms of the secular function:
\begin{equation}
Z_{scl}(E;\hbar) ={\rm det}\left (I- S(E,\hbar) \right ) \ ,
\label{eq:sclsecular}
\end{equation}
and the spectrum $\{ E_n \}$ satisfies the quantization condition
\begin{equation}
Z_{scl}(E_n;\hbar) =0\ .
\label{eq:sclcond}
\end{equation}
The secular function is nothing but the characteristic polynomial of $S$, 
$ {\rm det}  (I-z S (E,\hbar)) = \sum_{l=0}^L a_l(E,\hbar)z^l$ 
computed at $z=1$. The special symmetry of the
Riemann-Siegel expression, which is a consequence of the functional
equation in the Riemann-$\zeta$ function case, follows here from the 
unitarity of $S$.  It implies the inversive identities
 \begin{equation}
 a_l = e^{i\Theta}a^\ast_{L-l} \  ,
\label {eq:inversive1}
\end{equation}
 where  $\Theta $ stands for the phase of $\det\left (- S(E,\hbar)\right)$.
When the semiclassical approximation is used to compute $Z_{scl}(E;\hbar)$, 
the coefficients $a_l$ are expressed in terms of periodic orbits of the 
Poincar{\'e} map with periods up to $l$. Thus, the
computation of the semiclassical secular equation requires only a finite
number of periodic orbits.

 This proceedure was used e.g. by Bogomolny \cite {Bogomol} who derived a
semiclassical approximation for the restriction of the Green function 
to the Poincar{\'e} section, and by Doron and Smilansky  \cite {DoronUS} 
who used the scattering approach, and identified the scattering
matrix as the unitary operator in equation (\ref {eq:sclsecular}). Further
work along this line was carried out by Gutzwiller \cite {Gutzscat},  
Prosen \cite {prosen} and others 
(for a review see e.g., \cite{USLeshouches}).

The reduction of the classical dynamics to a discrete, area preserving
mapping, can also be achieved by observing the flow at fixed time 
intervals.  It is our aim here to suggest a new quantization scheme 
which is based on the quantum mechanical analogue of the stroboscopic
classical map.  The resulting quantization scheme, which is developed in
section (\ref{section:strobosc}), addresses the spectrum in a 
finite spectral interval of a size which is dictated
by the dynamical system and by Planck's constant. It is similar in its
formal structure to the 
methods which were briefly summerized above. However, it is based on the
classical periodic orbits of a different map, and it offers both the 
leading and the next to leading order contributions in the
semiclassical expression for the secular equation.  The semiclassical
version of the stroboscopic
quantization is given in (\ref {section:strobosc}). Among the other
consequences of our derivation
we note a relation between the time steps in the evolution operator and
the mean density of
states,  a consistent truncation of the periodic orbit series as well as
conditions  for
self-inversiveness of the characteristic polynominal.

The idea of extracting eigenvalues within a certain energy band has
many connections to frequently employed numerical schemes that
also obtain eigenvalues within a certain frequency band. For instance,
from the overlap of a wave packet with its propagated image over some time
interval one can extract eigenvalues in a band near the initial energy
\cite{Heller93}.
The semiclassical procedure described here achieves the same without
a weighting by the projection of initial wave packet onto eigenstates
and moreover suggests a semiclassical approximation. A numerically
accurate and efficient way of implementing this Fourier analysis has
been developped by Neuheuser and Wall \cite{Neu90,Wal95} and 
Mandelshtam and Taylor \cite{Man97a,Man97b,Man98} in the form 
of harmonic analysis. The way harmonic analysis beats the resolution
limit set by the Nyquist theorem is by implementing additional information
on the Fourier signal. In the case of the traces it uses explicitely
that they are sums of exponentials with a discrete set of frequencies.
We will relate the basic ideas of harmonic analysis to the semiclassical
traces and discuss the relation to the cycle expansion.
The relation of the present formalism and
harmonic inversion will be discussed in section (\ref {section:HI}). A few
concluding remarks and
a summary  will be the contents of section (\ref {section:conclusion}).

\section {Stroboscopic quantization} 
\label {section:strobosc}

We would like to compute a {\it finite} sequence of energy eigenvalues
in the spectrum of an autonomous (time independent) quantum Hamiltonian
$H$. This sequence is located in an energy interval
$ {\cal E}$,
\begin{equation}
 {\cal E} \equiv \left [E_0 -\ \Delta/2\ ,\   E_0 +\ \Delta /2\ \right ]\,.
\label{eq:domain}
\end{equation}
The midpoint of the interval $E_0$ is arbitrary, and its length
$\Delta $ is
$\approx {\cal O}(\hbar^{\gamma}) $ where $1> \gamma >0$. This makes $\Delta$
small on the classical scale but large
on the quantum scale since for a system with $f$ degrees of freedom
the mean spectral density $\bar d(E_0) \approx {\cal O}(\hbar^{f}) $. 
Hence the  number of eigenenergies in the interval
\begin{equation}
N =\bar d(E_0) \ \Delta \,,
\end{equation}
is large. Moreover,
$\Delta $ is also  sufficiently small so that  $\bar d(E)$ is effectively
constant in the interval.

Consider now the $N$ dimensional Hilbert space spanned by the 
eigenvectors of $H$ with
eigenenergies in ${\cal E}$. The quantum evolution operator in this
subspace can be
projected from the full evolution operator
$U(t)= {\rm e}^{-{i}Ht/\hbar }$. It is given explicitly by
\begin{eqnarray}
{\cal U}(t ; {\cal E}) \  &= &  \ \sum_{ E_n\in {\cal E} }
|\psi _n\rangle \, {\rm e}^{-{i} E_n t /\hbar }\, \langle\psi _n|
\label {eq:ucal}
\end{eqnarray}
and
\begin{eqnarray}
{\tr \cal U}(t ; {\cal E}) \  &= &  \ \sum_{ E_n\in {\cal E}}
 {\rm e}^{-{i} E_n t/\hbar  }
\label {eq:trucal} % \\
= \int \chi(E)\ d(E) {\rm e}^{-{i}  E t /\hbar} {\rm d }E
\,,
\end{eqnarray}
where $\chi(E)$ is the characteristic function of ${\cal E}$, and
\begin{equation}
d(E) = \sum_{\nu=1}^{\infty} \delta (E-E_\nu)
\label{eq:G}
\end{equation}
is the density of states for the full system. We will use small greek
indices to run over all eigenstates of the Hamiltonian and small
latin ones if they are restricted to the energy interval $\cal E$.

Choose an as yet arbitrary time interval of length $\tau$ so that
$\tau\Delta \le 2\pi \hbar$. 
The $N$ eigenvalues ${\rm e}^{-{i} E_n \tau /\hbar }$ 
of the unitary matrix ${\cal U}(\tau ; {\cal E}) $ occupy an arc of 
the unit circle whose length is at most $2 \pi$, and therefore they 
preserve the order of the spectral sequence on ${\cal E}$. The  
spectrum of  $H$ in the interval  
${\cal E}$ can be identified as the zeros of the secular equation: 
\begin{equation} 
\label{eq:secular} 
\zeta_{{\cal E}}(E) =  \det \left(  I- {\rm e}^{{i\over \hbar} E \tau }\ 
{\cal U}(\tau \ ; {\cal E})\  \right ) \,.
\end{equation} 
In other words, whenever the eigenphases of 
${\rm e}^{{i\over \hbar} E \tau }\ {\cal U}(\tau ; {\cal E})\ $ 
take the value $1$, a spectral point of $H$ is 
encountered.  Because of the exponentiation and the 
restriction to a finite set of eigenenergies in the interval 
$\cal E$, the density of states derived from this secular equation  
is a train of delta functions at the positions of the eigenenergies 
$E_n$, periodically continued by a separation $\hbar/2\pi\tau$, 
\begin{eqnarray} 
\label {eq:density} 
d(E;{\cal E}) &=&  \tau/  \hbar \ 
\sum_{m=-\infty}^{\infty} 
\sum_{n: \left \{ E_n\in {\cal E}\right\} } 
\delta (\ \tau(E-E_n )/\hbar + 2\pi m)   \nonumber\cr 
&=&{ N \tau\over  2\pi  \hbar}  + \  {\tau\over  2\pi  \hbar}\ \ 
\sum_{m=1}^{\infty}\left\{ {\rm e}^{-i m E \tau/\hbar}\ \  
\tr {\cal U}^m( \tau; {\cal E})  + c.c. \right\}\nonumber\cr 
&=&{ N \tau\over  2\pi  \hbar}  + \  {\tau\over  2\pi  \hbar}\ \ 
\sum_{m=1}^{\infty}\left \{ {\rm e}^{-i m E \tau/\hbar}\ \  
\tr {\cal U}( m\tau; {\cal E})  + c.c. \right \} 
\ . 
%\nonumber
\label{eq:trace_pol}
\end{eqnarray} 
The second line follows from Poisson summation of the first line. 
As usual, the density of states can be split into a smooth part and 
an oscillatory part, where the latter should vanish when 
averaged over the energy interval $\hbar/2\pi\tau$. 
In the semiclassical limit, the traces
$\tr
{\cal U}^m(\tau; {\cal E})$ are expressed as sums over periodic orbits, which 
give the {\it oscillatory} part of the spectral density. The smooth  
parts of both sides of (\ref{eq:density}) must be identical, which  
defines $\tau$ uniquely as 
\begin{eqnarray} 
 \tau = 
{2\pi\hbar {\bar d} (E_0) \over N }\ \ \ = {\tau_H \over N} \, , 
\label{eq:tau} 
\end{eqnarray} 
where $\tau_H$ is the Heisenberg time. 
 
This choice of the value of $\tau$ completely determines the quantum map 
${\cal U} ( \tau; {\cal E})$ which provides us through (\ref {eq:density}) 
with the spectral density of $H$ in ${\cal E}$. This is a  
{\it stroboscopic} map, since $\tau$ is fixed. The unitarity of  
${\cal U} ( \tau; {\cal E})$ is an important asset, since one can  
write the secular equation in a form which exploits naturally the  
inversive symmetry of the coefficients of the characteristic 
polynomial of ${\cal U} ( \tau; {\cal E})$. Writing 
\begin{equation} 
\zeta_{\,{\cal E}}(E) =   \det \left(  I- \ {\rm e}^{{i} E \tau/\hbar 
}\  {\cal U}(\tau ; {\cal E})\  \right )\   = \sum_{n=0}^N a_n 
{\rm e}^{i n E \tau/\hbar }\,,  
\label {eq:polynom} 
\end{equation} 
the inversive symmetry is expressed by the identity 
 \begin{equation} 
 a_n = e^{i\Theta}a^\ast_{N-n} \  , 
\label {eq:inversive} 
\end{equation} 
 where  $\Theta $ stands for the phase of $\det (- {\cal U}(\tau ; {\cal 
E}))$. 
We use 
\begin{equation} 
\det \left ( 
{- \rm e}^{i E \tau/\hbar }\  {\cal U}(\tau ; {\cal E}) \right)= 
{\rm e} ^ {-i \pi N + i 2\pi (E-E_0){\bar d}(E_0)}  =  {\rm e} ^{i 2\pi 
{\bar 
N(E,E_0)}}\ , 
\end{equation} 
where we introduced the smooth spectral counting function in the energy 
interval ${\cal E}$, ${\bar N}(E,E_0)=(E-E_0)\bar d(E_0) - N/2$. Thus, 
\begin{eqnarray} 
{\rm e} ^{i \pi\ {\bar N(E,E_0)}}\zeta_{\ {\cal E}}(E) = 
\sum_{n=0}^{[N/2]}\left \{ 
a_n{\rm e}^{ i\pi\ {\bar N(E,E_0)}} +a^*_n{\rm e}^{-i\pi\ {\bar 
N(E,E_0)}}\right \} 
+\epsilon_{N} a_{N/2}{\rm e}^{ i\pi\ {\bar N(E,E_0)}}\,. 
\label {eq:finalzeta} 
\end{eqnarray} 
As before, $[x]$ stands for the largest integer smaller than $x $, and 
$\epsilon_{N} =1$ if $N$ is even and $\epsilon_{N} =0$ otherwise. 

The cycle expansion\cite{Pl,Sm,RS} and Newton's identities 
 relate the traces  
$t_n = \tr  {\cal U}^n(\tau; {\cal  E})$ 
 and the coefficients of the characteristic polynomial 
 \begin{equation} 
 a_n = -{1\over n} \left(t_n +\sum_{k=1}^{n-1}a_k t_{n-k} \right ) \ . 
\label {eq:Newton} 
\end{equation} 
 The inversive symmetry halves the number of coefficients needed to 
calculate the spectral secular equation. This symmetry is a  
consequence of the unitarity of the evolution operator and  
as such it is the expression of a basic property of the 
quantum evolution - the conservation of probability (norm). The semiclassical 
theory provides approximate expressions to $t_n$ which are used to compute 
the $a_n$ using Newton's identities. The resulting semiclassical  
secular function is similar  
in many ways to the expressions derived by  
Berry and Keating \cite{BK90,BK92,K92}, Bogomolny \cite{Bogomol}, and  
Doron and Smilansky \cite{DoronUS,USLeshouches}.  
It will be discussed in the sequel. 
 
\section {The semiclassical approximation}  
\label {section:scl} 
 
It is convenient to introduce the semiclassical approximation by writing 
 (\ref {eq:trucal}) as 
\begin{eqnarray} 
 \ \tr {\cal U}(t;{\cal E} ) 
&=& %%%%{1\over 2\pi \hbar }  
\int \chi(E) d(E) {\rm e}^{-{i}  E t/\hbar } 
{\rm d }E 
 \nonumber \\ 
& = & {1\over 2\pi \hbar } \int_{-\infty}^{\infty}{\rm d } s 
\ \hat \chi(s) \ \tr U(t+s) \ , 
\label{eq:uu} 
\end{eqnarray} 
where 
\begin{equation} 
 \hat \chi(s) = \int_{-\infty}^{\infty} \ \chi(E) {\rm e}^{ {i\over \hbar }E 
s}{\rm d}E \  =\  {\rm e}^{ {i}E_0 s/\hbar} \ 
 {\sin {s\Delta\over 2  \hbar } \over { s\over 2  \hbar } } \ , 
\end{equation} 
and where $U(t)= {\rm e} ^{-{i}Ht/\hbar} $ is the evolution operator of the 
{\it entire } system. We can then use the 
semiclassical expression for $\tr U(t )$ for any time $t$, given by 
\begin{eqnarray} 
 \left [ \tr U (t) \right ]_{scl} 
\ =\  {t \over  (2\pi i \hbar )^{1/2}} \ \sum_{p\in {\cal P} (t )} 
\left ( {{\rm d}E_p\over {\rm d} t}\right )^{1/2} {{\rm e} ^{i R_p/\hbar 
-i\nu_p{\pi/ 2}} \over \left | \det(I-M_p) \right |^{1/2}} \,, 
\label{eq:scltrace} 
\end{eqnarray} 
where 
${\cal P} (t)  $ is the set of $t$-periodic orbits  and 
\begin{eqnarray} 
R_p = \int _0 ^{t} \left [\ p_p(t) \dot q_p(t) - H\left (q_p(t),p_p(t)\right 
)\  \right ] 
{\rm d} t \ . 
\label{eq:action} 
\end{eqnarray} 

The building blocks for the semiclassical theory are the traces
$ \tr {\cal U}^n(\tau;{\cal E})$. By an exact quantum 
identity they coincide with $\tr {\cal U}(n \tau;{\cal E})$, 
so that we need
to compute the traces for times which are integer multiples of $\tau$. 
Notice that the periodic orbits $p\in {\cal P} (t)  $ which contribute to 
(\ref {eq:scltrace}) can have {\it arbitrary energies} 
$E_p(t)$. They are functions of the period $t=n\tau$ which is {\it fixed}. 
Substituting (\ref {eq:scltrace}) in  (\ref {eq:uu}), 
\begin{eqnarray} 
 \left [ \tr {\cal U}(n\tau;{\cal E} ) \right ]_{scl} &=& {1\over 2\pi 
\hbar } 
\int_{-\infty}^{\infty}{\rm d}s \ \hat \chi(s) \ \left [ \tr 
U(n\tau+s) \right]_{scl} \nonumber \\ 
&\approx & {1\over 2\pi \hbar } 
\int_{-\infty}^{\infty}{\rm 
d}s \ \hat \chi(s) \ {n\tau+s\over  (2\pi i \hbar )^{1/2}} \ \sum_{p\in {\cal 
P} (n\tau+s )} 
\left ( {{\rm d}E_p\over {\rm d} t}\right )^{1/2} {{\rm e} ^{i R_p/\hbar 
-i\nu_p{\pi/2}} \over \left | \det(I-M_p) \right |^{1/2}} \,.
\label{eq:scltruu} 
\end{eqnarray} 
The range of $\hat \chi(s)$ is determined by the time 
${\hbar \over \Delta }\approx {\cal O}(\hbar^{1-\gamma }) $ 
which is  classically short. Hence we can approximate the integral over the 
periodic 
orbit sum by expanding the phase of the integrand about $t$. For this 
purpose we 
use ${\partial R_p(t) \over \partial t} = - E_p(t)$.  Denoting the 
preexponential amplitudes in  (\ref {eq:scltruu}) by $A_p(t) $, 
we use $A_{p}(n\tau+s)\approx A_{p}(n\tau)$  which is consistent with the 
semiclassical approximation.  The contribution of 
each periodic orbit will be 
\begin{eqnarray} 
&&{ A_{p}(n\tau)\over 2\pi \hbar }\ {\rm e}^{{i\over \hbar } S_p(n\tau)} 
\int_{-\infty}^{\infty}{\rm d}s \ {\sin {s \Delta \over 2  \hbar } \over { 
s\over 2 
\hbar } }\ {\rm e}^{{i\over \hbar }\left 
[(E_0-E_p(t))s -{1\over 2} {{\rm d}E_p\over {\rm d} t}s^2\right ]}\nonumber 
\\ 
&=& A_{p}(n\tau)\ {\rm e}^{{i\over \hbar } S_p(n\tau)} \left ( {-i\over 2} 
\right 
)^{1\over 2} 
 \left \{ \left [ C(x^+_p(n\tau) ) + i S(x^+_p(n\tau)) \right ] 
- \left [ C(x^-_p(n\tau) ) + i S(x^-_p(n\tau)) \right ] \right \}\ , 
\label{eq:sclamplit} 
\end{eqnarray} 
where $C$ and $S$ are the $ \cos$ and $\sin$ Fresnel integrals with the 
argument 
\begin{equation} 
x^{\pm}_p(n\tau) =  {E_p(n\tau) - E _0 \pm {\Delta \over 2} \over \left(2 \hbar 
\left |{{\rm 
d}E_p\over {\rm d} t}\right |\right)^{1\over 2} } 
\label{eq:Fresnel} 
\end{equation} 
In the above expression, the difference of the energy of the periodic orbit, 
$E_p(n\tau)$, from the interval of interest ${\cal E}$, is measured in 
units of the energy 
scale $(2 \hbar \left |{{\rm d}E_p\over {\rm d} t}\right |)^{1\over 2}$. 
(we assumed throughout the derivation that, as is typically the case, 
${{\rm d}E_p\over {\rm d} t}$ is negative.) 
Using dimensional analysis one can bound this scale from above by 
${\cal O} \left (\hbar^{{1-\gamma \over 2}}\right )$, which is small on the 
classical scale. Using the asymptotic expression for the Fresnel integral 
for large argument, 
\begin{equation} 
C(x) +iS(x) \approx \left ({i\over 2} \right )^{1\over 2}{\rm sign}(x) - 
{i\over x 
\pi}{\rm e}^{i{\pi\over 2}x^2} + {\cal O} \left ({1\over |x|^3}\right ) \,. 
\label {eq:asympfres} 
\end{equation} 
The difference between the Fresnel integrals in (\ref {eq:sclamplit})  
is dominated in the semiclassical limit 
by the contribution of the leading term 
in (\ref{eq:asympfres}) which, 
together with the factor  $\left ({-i/ 2} \right )^{1/ 2}$ (see 
(\ref {eq:sclamplit})) is the characteristic function of the spectral 
interval ${\cal E}$. Thus,  the  leading  semiclassical contribution
makes use of periodic orbits whose energy 
$E_p(t) $ is in ${\cal E}$ and does not take into account the 
``diffractive" 
effects due to the sharp restriction of the spectrum to the interval 
${\cal E}$. The 
Fresnel functions with  finite arguments include the appropriate corrections. 
 
The  semiclassical approximation for the  $\left [t_n\right ]_{scl} = \left 
[ \tr {\cal U}(n \tau;{\cal E})\right ]_{scl}$ is obtained by summing the 
amplitudes (\ref{eq:sclamplit}) over the set  
${\cal P}(n \tau )$ of periodic orbits. When 
$\left 
[t_n\right ]_{scl}$ is substituted in (\ref {eq:finalzeta}) one obtains the 
semiclassical spectral secular equation. It can be written as a sum of two 
parts. The 
first is obtained from the leading semiclassical expression for 
$t_n$. It uses the periodic classical orbits of period $n\tau$ 
with energies 
in the  interval of interest ${\cal E}$. The diffractive corrections go 
beyond this 
limit, and apart from modifying the contributions of ``allowed" periodic 
orbits, it introduce the effects of periodic orbits whose energies are 
outside the 
strict energy interval. These corrections are analogous to the expressions 
derived by 
Berry and Keating \cite{BK90,BK92,K92}  
which are missing in the derivations based on the 
Poincar{\'e} 
section \cite{Bogomol} or the scattering approaches to quantization 
\cite{DoronUS,USLeshouches}. Here, one can 
attribute them to 
the sharp truncation of the energy domain and their main effect is the 
inclusion of evanescent, classically forbidden contributions. 

Finally, we would like to mention that the semiclassical approximation
could be introduced in a different way. With 
\beq
P({\cal E}) = \sum_{n:\{E_n\in{\cal E}\}} | n \rangle\langle n|\,,
\eeq
the projector onto the energy interval, we have the quantum identity
\beq
\tr{\cal U}^n(\tau;{\cal E})=\tr{\cal U}(n\tau;{\cal E})
=\tr\left( P({\cal E}) U(\tau)\right)^n \,.
\eeq
The semiclassical approximation we used in (\ref{eq:scltrace}) and 
(\ref{eq:scltruu}) is
\beq
\tr{\cal U}^n(\tau) \sim \left\lbrack
\tr {\cal U}(n\tau)\right\rbrack_{scl}\,.
\label{favor}
\eeq
The alternative approximation
\beq
\tr{\cal U}^n(\tau) \sim \tr\left( \left\lbrack 
P({\cal E}){U}(\tau)\right\rbrack_{scl}\right)^n\,.
\eeq
would contain a product of Fresnel factors that reduces to (\ref{eq:scltruu})
only to leading order. Such differences are not uncommon in attempts
to go beyond the leading order in the semiclassical approximation. We
favor the representation (\ref{favor}) since it makes use of as much
exact quantum information as possible and leaves the semiclassical
approximation only to the very end.
 
\section {Obtaining eigenvalues}  
\label {section:HI} 
 
Given the traces of ${\cal U}(n\tau)$ at equidistant time intervals 
$\tau$, the extraction of eigenvalues becomes a problem in Fourier 
inversion. If all traces $\tr{\cal U}^m$ in (\ref{eq:trace_pol}) and
(\ref{eq:Newton}) are known, the relation
is exact. However, in practice only a finite number of traces
can be calculated, and then standard Fourier inversion  
is limited in resolution by the Nyquist sampling theorem, which for
traces up to $N$ implies a resolution of the order of the 
mean spacing. Additional errors are introduced by the semiclassical
approximation. In order to go beyond the limit of resolution
set by the Nyquist theorem additional information has to be
added to the Fourier inversion. For the case at hand, harmonic
inversion \cite{Neu90,Wal95,Man97a,Man97b,Man98,main} is particularly
appropriate. It assumes that the signal is composed of a finite
number of frequencies and provides an efficient method for solving
the ensuing approximation problem. Specifically, it uses the 
traces to set up a matrix $\tilde V$ that has the same 
eigenvalues as ${\cal U}$,
so that the frequencies can be found from the traces by an
eigenvalue determination, which numerically is more reliable than
a standard search for zeroes. Note that the matrix $\tilde V$ has the 
same eigenvalues as ${\cal U}$, but it will usually not be unitary.
In the description of harmonic analysis below we will emphasise the
matrix structure and its origin and will omit numerical issues 
such as windowing and the like.
 
Let $N$ be the fixed dimension of the matrix ${\cal U}$ and let, as before,
\beq 
t_m = \tr {\cal U}^m = \sum_{n=1}^N e^{-im \phi_n}   \,,
\label{eq:traceform}
\eeq 
be the traces of the $m$-th power of the unitary operator. 
In the quantum case the phases $\phi_n=E_n \tau /\hbar$ are real and
contain the eigenenergies. In the semiclassical approximation they may
become complex, in which case the imaginary parts can be used
as a measure of the semiclassical error.
The form (\ref{eq:traceform}) of the traces implies that
$t_n$ for $n>N$ can be expressed as a linear combination of the traces
$t_m$ with indices $m<N$, as in the 
case of autoregressive models. Harmonic inversion exploits
these relations for the construction of a matrix $\tilde V$ that
has the same eigenvalues as ${\cal U}$.
To derive this matrix, define an $N$-dimensional vector ${\bf t}_m$ of 
traces starting with $t_m$, i.e., 
\beq 
{\bf t}_m = \left(\matrix{ t_m \cr \vdots \cr t_{m+N-1}}\right)\,, 
\eeq 
the $N$-dimensional vector ${\bf I} = (1,\cdots,1)^T$ 
and the auxiliary $N\times N$ matrix 
\beq 
\Omega_{m,n} = e^{-im\phi_n}  \,. 
\eeq 
Then 
\beq 
{\bf t}_1 = \Omega\, {\bf I} \,. 
\label{t1} 
\eeq 
For higher $m$ the vectors ${\bf t}_m$ can be represented as 
\beq 
{\bf t}_m = \Omega V^{m-1} {\bf I} \,. 
\label{tn} 
\eeq 
with the diagonal matrix 
\beq 
V_{m,n} = e^{-i\phi_n} \delta_{m,n}\,. 
\eeq 
The aim is to find a matrix $\tilde V$ that is similar
to $V$, so
that the frequencies can be determined from the 
eigenvalues of $\tilde V$. To this end solve (\ref{t1}) for 
${\bf I}$ and substitute in (\ref{tn}): 
\beq 
{\bf t}_m = \Omega V^{m-1} \Omega^{-1} {\bf t}_1 \,. 
\eeq 
Since also 
\bea 
{\bf t}_{m+k} &=& \Omega V^{m-1}\, V^k \Omega^{-1} {\bf t}_1 
= \left(\Omega V^{m-1} \Omega^{-1}\right) \,  
\left(\Omega V^k \Omega^{-1} {\bf t}_1 \right)\cr 
&=& \Omega V^{m-1} \Omega^{-1} \, {\bf t}_{k+1} \,, 
\eea 
it is possible to construct the full matrix 
\beq 
\tilde V_{m-1} = \Omega V^{m-1} \Omega^{-1} 
\eeq 
from the images of the vectors ${\bf t}_n$ for different $n$. 
Specifically, for $m=2$, the entries for the first iterate of
$\tilde V$ are 
\beq 
{\bf t}_{k+2} = \tilde V_1 {\bf t}_{k+1}\,. 
\label{trace_relation} 
\eeq 
This result shows that the matrix $\tilde V_1$ can be 
constructed from the traces (assuming that the eigenvalues 
are not identical and the vectors of traces not linearly dependent).  
The matrix $\tilde V_1$ is 
neither unitary nor symmetric but nevertheless has 
the eigenvalues $\exp(-i\phi_n)$. 

It is possible to proceed one step further and 
to derive the characteristic polynominal. 
With the $N\times N$ matrices 
\beq 
T_{M} = 
\left(\matrix{t_{M} & t_{M-1} & \cdots & t_{M-N+1} \cr 
           t_{M+1} & t_{M} & \cdots & t_{M-N+2} \cr 
           \vdots & \vdots & & \vdots \cr 
           t_{M+N-1} & t_{M+N-2} & \cdots & t_{M}} \right) \,,
\eeq 
the relation (\ref{trace_relation}) becomes 
\beq 
T_{N+1} = \tilde V_1 T_N \,, 
\eeq 
so that formally $\tilde V_1=T_N (T_{N+1})^{-1}$. 
Since $N-1$ columns 
in $T_{N+1}$ and $T_N$ coincide (up to a shift to the right), 
$\tilde V_1$ has the form 
\beq 
\tilde V_1 = \left(\matrix{0 & 1 & 0 & 0 & \cdots & 0 \cr 
                    0 & 0 & 1 & 0 & \cdots & 0 \cr 
                  0 & 0 & 0 & 1 & \cdots & 0 \cr 
                 \vdots & \vdots & \vdots &\vdots  &  & \vdots \cr 
             b_{N} & b_{N-1} & b_{N-2} & b_{N-3} & \cdots & b_1} \right) 
\,,
\eeq
where the vector ${\bf b}=(b_N, b_{N-1}, \cdots, b_1)$ solves 
\beq 
\left(\matrix{t_{2N-1}\cr t_{2N-2} \cr \vdots \cr t_{N-1}} \right) 
= T_N^T 
 \left(\matrix{a_{N}\cr a_{N-1} \cr \vdots \cr a_{1}} \right) 
\,.
\label{coeff_eq}
\eeq 
Exploiting this special form of the matrix $\tilde V_1$ and the 
vector ${\bf b}$ the Fredholm determinant can be expanded 
\beq 
F(z) = \det(1-z\tilde V_1) =  1 - \sum_{k=1}^N z^k b_{k}\,. 
\label{eq:Fharmonic} 
\eeq 
Since the eigenvalues of ${\cal U}$ and $\tilde V_1$ are 
the same, the coefficients of the characteristic polynominals 
(\ref{eq:polynom}) and (\ref{eq:Fharmonic}) also have to be the same, 
up to an overall scale factor. 
 
It is important to note one significant difference, though: 
in the derivation of the polynominal for the quantum operator 
${\cal U}$, traces up to order $N$ determined all coefficients
and self inversivness allowed to bring this number down to $N/2$. 
In the harmonic inversion case the number of traces needed is 
$2N+1$. The origin of this difference is the fact that in the 
case of the harmonic inversion nothing changes in the 
formalism if the traces $\tr {\cal U}^n$ are replaced by 
$\tr A {\cal U}^n$ with some operator $A$: the set up for the matrix 
$\tilde V_1$ remains the same, but the coefficients of the 
vector ${\bf I}$ change. If $A=1$, then traces 
with $n>N$ can be expressed by traces of lower order 
using Cayley's theorem. Thus, harmonic inversion is 
more general in its basic structure, but also more 
wasteful in the number of traces required. In addition, 
it is difficult to see how properties such as  
self-inversiveness can be implemented directly
(they can always be put in by hand in the
characteristic polynominal). 
 
\section {Concluding remarks}  
\label {section:conclusion} 
 
The main features of stroboscopic quantization are the 
limitation to a finite interval in energy, the semiclassical 
expression with a finite number of periodic orbits and 
the finite characteristic polynominal. The origin of 
the finite representation is a physical one, the 
limitation to a finite interval in energy. Other modifications 
of the Gutzwiller trace formula, such as a Gaussian 
truncation or even a sharp cut-off, arrive at this 
restriction at the price of smearing the eigenvalues. 
Moreover, since with these smearings the evolution operator 
is not confined to a finite interval, the characteristic 
function is not a polynominal anymore and the conditions of  
self-inversiveness are difficult to implement. 
 
The weight of periodic orbits and in particular the 
Fresnel corrections bear a striking similarity to the 
error function truncations introduced by Berry 
and Keating on the level of the Fredholm product \cite{BK90,BK92,K92}.  
Note, however, that the product of two Fresnel integrals is 
not a Fresnel integral of the sum of the arguments, 
so that the modification that would apply to 
a pseudo-orbit differs from the product of the weights 
of the original orbits. 
 
On the numerical side, the results on stroboscopic 
quantization suggest that for harmonic analysis 
the time period $\tau$ over which the evolution operator 
is followed should be fixed to be $\tau_H/N$, so that 
an optimal characteristic polynominal that 
shows self-inversiveness results. Ideally, 
one would like to improve the method so as to 
use only the linearly independent $N/2$ traces, but 
it is not clear how that can be implemented other
than by brute force.
 
The finiteness of the characteristic polynominal 
results from the quantum fact that after the projection onto 
the energy interval the Hilbert space is finite dimensional 
and that there are only a finite number of eigenvalues. 
In particular, this implies relations between 
traces of higher powers of ${\cal U}$ due to  
Cayley's theorem. Within the semiclassical approximation 
no such constraint and no such relation between 
traces of higher and lower powers are evident. 
In many approaches \cite{BK90,BK92,K92,Bogomol,DoronUS,USLeshouches}  
they are put in  by hand and justified by appeal to the properties 
of the quantum propagator. However, it is legitimate 
to ask to which extend the semiclassical approximation 
reflects these properties by itself, and it is 
comforting to note that, as demonstrated 
for the Bakers map \cite{EH} and also for the three  
disk scattering system \cite{Pinball,Wirzba,ER}, where none of these 
unitarity arguments apply, the coefficients 
in the cycle expansion beyond the dimension of the 
system decay.  
 
\section*{Acknowledgements}
BE would like to thank VA Mandelshtam for a stimulating discussion
on harmonic inversion.
US thanks the Alexander von Humboldt foundation for support
and the Philipps Universit\"at for its hospitality. This work
was also supported in part by the Minerva Center for Nonlinear Physics 
at the Weizmann Institute of Science.

%%%%%%%%%%%%%%%%%%%%% 
 
\end{document}